# What Clinical Trials Can Teach Us about the Development of More Resilient AI for Cybersecurity

Edmon Begoli, Robert A. Bridges, Sean Oesch, Kathryn E. Knight

Oak Ridge National Laboratory (ORNL), Oak Ridge, Tennessee, USA

Key Words: artificial intelligence, resilient AI, cybersecurity policy, clinical trials, evaluation and test

## Abstract

Policy-mandated, rigorously administered scientific testing is needed to provide transparency into the efficacy of artificial intelligence-–based (AI-based) cyber defense tools for consumers and to prioritize future research and development. In this article, we propose a model that is informed by our experience, urged forward by massive scale cyberattacks, and inspired by parallel developments in the biomedical field and the unprecedentedly fast development of new vaccines to combat global pathogens.

## Introduction

As human health is essential to a functioning society, vaccines have proven critically important for dampening and even thwarting many diseases that have plagued humanity (MacDonald et al., 2020). Yet, to be used, vaccines must be endorsed by regulatory bodies and are used only if they pass a set of rigorous, staged scientific trials aimed at (1) guaranteeing a large benefit-to-detriment ratio in use and (2) providing visibility into at least some scientifically grounded measures of efficacy for the public. By analogy, we argue that the world's economy, critical infrastructure, society, and culture depend on healthy, reliable, and functional networks, which in turn are reliant on computer network defense technologies (e.g., malware detection, network intrusion detection) as critical defenses from cybercriminals, state-sponsored actors, and other adversarially minded groups. Yet, the scale and the size of the problem transcends any single tool or approach, requiring artificial intelligence–based (AI-based) approaches to help us cope with the scope, scale, complexity, and uncertainty of the problem. However, no policy for mandating minimal effectiveness, reliability, and transparency of these defensive, especially AI-based, measures exists. Enacting such policy is a gargantuan undertaking, and we recognize that fact. In this paper, we take a proverbial bird's eye view of the well-established policies in critical domains, such as the vaccination development process, and zoom in to discuss organically occurring analogous testing of commercial-off-the-shelf cyber tools currently underway. This provides a real-world example of how cyber analogs to clinical trial policies may strengthen cyber defense and structural efforts for future developments. The promises of AI and machine learning (ML) in computer defense, in particular, are considered.

This discussion is particularly relevant in light of the recent SolarWinds attack (Tung, 2021), which was reported to have impacted over 250 US federal agencies and businesses (Schneier, 2021). In response to this attack, security experts have highlighted the need to "improve government software procurement" (Schneier, 2021). The authors' research organization, Oak Ridge National Laboratory, is currently aiding US federal agencies in testing intrusion detection technologies that are powered by AI/ML to ensure that the technologies selected are maximally effective (McDermott, 2020). Admittedly, the authors are not policy experts, nor do we have the necessary data to make detailed policy recommendations. However, we are reaching into our unique experiences with these novel technologies with the hope of informing future best practices and to spur conversations that can lead to the development of more effective cybersecurity policies for certification of AI-based cyber defenses.

# Background

In this section, we provide a brief background on the clinical trial process and the role that AI plays in cybersecurity. Then, we will then make suggestions for how to map AI-based cyber tool development to the clinical trial process and exemplify how we are already using elements of the clinical trial process in practice to aid in the development of more reliable cyber tools. We will then discuss the implications of our experiences and what effective cybersecurity policy might look like in practice.

## Rigor in Clinical Trials

To introduce a new vaccine that demonstrably immunizes against the target pathogen(s), the proposed treatment needs to undergo a sequence of rigorous clinical trials providing staged evaluations and gradual introduction of the new treatment. These trials, in general, consist of three or four stages (Evans, 2010). In the first stage, the proposed vaccine is evaluated for the proposed validity of its design; subject matter experts evaluate the planned treatment to assess its merit. In the next stage (or a trial), the proposed vaccine is tested in a limited and controlled laboratory setting, often on lab animals. In the third stage, the proposed vaccine is tested with a limited patient population and in a double-blind protocol; that is, for an ideally large number of subjects, both the actual treatment and a placebo treatment are administered identically with both the recipients and those administering the treatment blind to who received real or placebo vaccines. Comparing the results of infection between the groups provides statistical tests to determine efficacy of the treatment in the real world. Once the efficacy and safety standards are reached, the treatment is approved for use in the general population, but the results of the application of the treatment in the population are monitored for safety, efficacy, and adverse effect, which could result in a recall of the vaccine. This is the fourth phase or stage of the clinical trial.

Artificial Intelligence Possibilities and Pitfalls

      Two of the major concerns in network defense are (1) the ability to detect attackers once they are in the network and (2) the speed at which a response occurs once they are detected. AI-based security solutions have the potential to solve both detection and response problems. Examples in the cybersecurity market abound. Decision classifiers (used in supervised learning) are already replacing malware detection at the endpoint and network levels, while more sophisticated algorithm suites are packaged to identify suspicious behaviors via network traffic and other logged data and are meant to complement rule-based systems. Driven by insider threat concerns, a submarket of User and Entity Behavior Analytics (UEBA) tools use ML-based models to detect anomalous user or machine behavior in the network, which could significantly reduce the time it takes to detect attackers once they are in the network. Security, Orchestration, Automation, and Response (SOAR) solutions promise to increase automation of triage and response, thereby potentially reducing response time.

      While AI offers unique capabilities that can help make computer networks more secure, it also increases the attack surface available to adversaries and presents new avenues for attackers to exploit (Liwel et al., 2019; Katzir & Elovici, 2018). More generally, there are numerous examples (Uesato, 2018) of AI vulnerabilities, in and outside of the cyber context. As recent research shows (Li, 2018; Song, 2019), all of these qualities are also subject to adversarial exploitation and can be a source of privacy and security vulnerabilities. For example, the models can be trained on purposefully and intentionally tampered training data. This approach is known as data poisoning. The models trained on such "poisoned" data would in effect make AI models develop a "blind spot" for the phenomena that adversaries are interested in exploiting once the model is deployed in operations. Another approach (i.e., adversarial perturbations, data patches) manipulates the input to the already-trained model with an intention to confuse it. The goal of this kind of exploitation is to confuse the defenses and have them mistake the malicious input for valid, therefore passing the defenses undetected. In summary, the statistical machinery and computing power of AI can be and is used to poison, fool, or otherwise thwart a second AI system. This emerging, fast-growing research area, called "adversarial AI," is continually producing new methods. That dynamic makes the need for a continuously updated, rigorous, formal, and principled approach for testing and evaluating reliable and resilient AI all that more important.

One view (discussed informally[1] and explored rigorously[2]) is that security, physical or otherwise, is the practice of introducing asymmetry by creating enough difficulty that exploitation is no longer worth the cost. Because AI is providing powerful automation tools to either instantiate or alleviate difficulties, it stands to benefit from both defensive and offensive actors. While it is still too early to tell whether attackers or defenders will benefit most from AI, an argument can be made that AI will ultimately tip the balance in favor of the defenders if it significantly reduces the time required to detect intruders once they enter the network (Buchanan, 2020). This possibility is highly relevant to policymakers at the government level because the offense–defense balance impacts strategic stability and may influence whether or not nation-states decide to go on the offensive (Jervis, 1978; Glaser, 1997). Overall, there is a need to make AI more resilient and reliable for it to be used in critical domains such as cyber defense.

## Taking Inspiration from the Structure of Clinical Trials

Here we consider the clinical trial stages and attempt a hypothetical mapping of their meta structure to the conceptual stages by which AI-based cyber defenses can be rigorously and methodically evaluated for resilience to adversarial exploitation and for effectiveness in large-scale network defense. We argue that the challenges of defense-based AI research are analogous to the scientific challenges posed by medical treatment research, where a pathogen may be either known or novel, with treatments often requiring continuous (re)evaluation and improvement. As such, we propose four phases of evaluation comparable to clinical trials, with each phase examining the test design elements and methodology, approaches to controlled and large-scale evaluation, and ongoing monitoring requirements, respectively.

In Phase 1, the focus is on the rigorous evaluation of methods and design characteristics. The objective of this phase is to explore the theoretical and conceptual aspects of the approach before the larger-scale tests are conducted. The attempt is to answer the question "*Will it work?*" before further testing is attempted. Only small-scale tests are conducted to evaluate the critical design and methodological aspects of the approach, including its fundamental efficacy and the soundness of the defensive AI models.

In Phase 2, the next stage of expanded evaluation happens in a controlled environment and against a larger number of malicious and benign software components. Some network components can be simulated. The focus of this phase is on the effectiveness, reliability, and possible side effects of the approach, including critical vulnerabilities or inability of AI to respond to known malware, and the well-known adversarial AI vectors.

---

[1] https://polyverse.com/blog/how-to-think-about-security-asymmetry-of-difficulty-498eeebe91b5/.
[2] L. Wang, A. Singhal, and S. Jajodia. (2007). "Measuring the overall security of network configurations using attack graphs." In the Proceedings of the IFIP Annual Conference on Data and Applications Security and Privacy. Springer, Berlin, Heidelberg, 2007.

In Phase 3, the large-scale evaluation begins. It happens in an operationalized-like environment, with common network, system, and software components in place, and against a large library of malware (as well as "benignware"). The testing in this phase in all respects mimics a production-like environment, and it is designed to statistically match the population and the distribution of threats found in the "open." The emphasis of this phase is on the efficacy of the AI solution in the function of cyber defense and the success ratio against any known and speculative configuration of threats.[3]

Phase 4 is a monitored deployment phase. The cyber defense solution is deployed to the production environment and operationalized. Following inspiration from clinical trials, the deployed defense is monitored continuously for safety and efficacy. Similar to the deployment of new therapies, the deployed AI-based tools can be recalled if they are observed to be vulnerable to adversarial exploitations, or that they do not function properly.

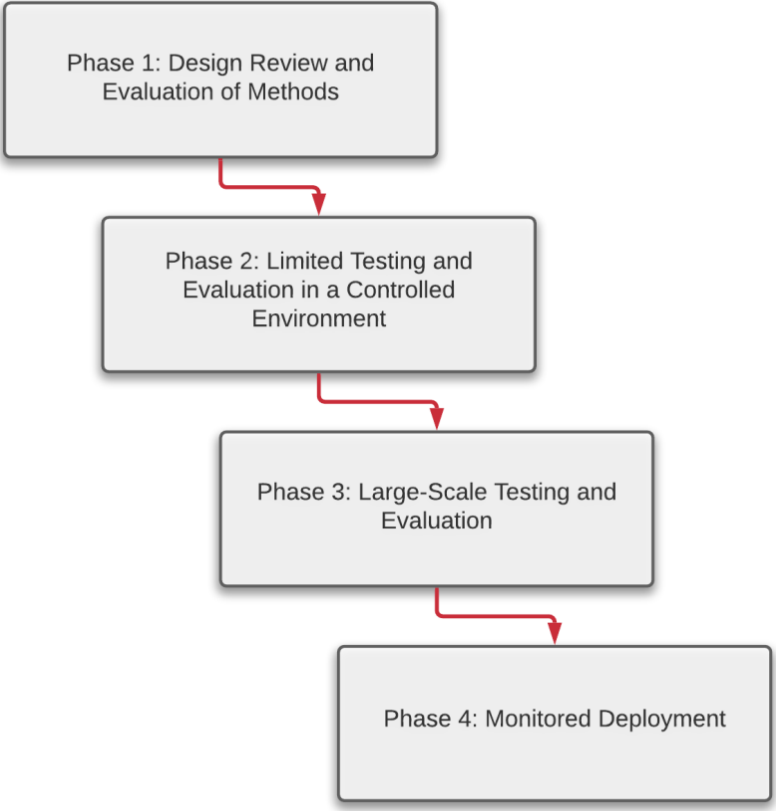

Figure 1. A proposed structure of evaluation phases.

**Emerging Lessons Derived from Practice**

The recommendations presented in this work are inspired by two principal sources. One is our experience in working on a similar but less rigorously structured testing and evaluation program for AI-based cyber defense tools. The other inspiration is our experience in AI use in biomedicine and drug discovery. Naturally, the new inspiration occurred at the intersection of these two domains.

---

[3] Threats such as complex, novel malware and advanced persistent attacks can be simulated using AI itself.

Related to AI-based cyber defenses, we are engaged in an ongoing effort to strengthen the computer network defense tools of research sponsors in what is currently a three-part process. Each step requires an ideally impartial set of evaluators with technical expertise and cooperation of the companies developing and/or selling technologies of interest. (As it happened, we had devised the evaluation process that, as we observed later, happened to map to the first three stages of a vaccine clinical trial.)

First, subject matter experts (e.g., account executives and engineers) from the companies or laboratories promoting a cybertechnology meet with an impartial third party (in this case, the research scientists) to convey and defend their technical approach. The goal is to allow the scientists to determine the merits based on the concepts put forth. Over time, this market research allows the scientists to see a wide variety of ideas and corresponding technologies in each subclass of the market (e.g., malware detection, UEBA, SOAR). Our experience with doing this type of market research with a few dozen companies is that while companies are generally proud to provide insights into the intellectual merits of their approach, the propensity for salesmanship clouds the merit of the discussion. This motivates the subsequent steps.

Secondly, an experiment is designed and carried out with a goal of deeply testing competing technologies head-to-head in a controlled laboratory environment. Medium-scale examples of such testing include tests as described in our concurrent work (Bridges et al.) where thousands of malware/benignware files of different types are used to evaluate endpoint and network-level detection abilities. Such a test provides great insight to consumers of those specific technologies and a glimpse into the state of the art. Examples of large-scale experiments in this vein include the AI ATAC Challenges (Naval Information Warfare Systems Command, US Department of Defense [NAVWARSYS], n.d.-a, n.d-b, n.d.-c) In the first, endpoint malware detection (commonly called antivirus, or AV) technologies were stress tested with 100K files in a completely automated, highly parallelized custom framework while many measurements (e.g., CPU, memory, bandwidth resource) were gathered. In the second, a full network with services (e.g., email, ssh), emulated and real Internet access, and emulated users, was designed and implemented to test network traffic analyzers' ability to detect multistage attack campaigns. Hundreds of adversarial campaigns were waged against the network to identify strengths and weaknesses of the detection tools. (The third AI ATAC Challenge is still under development.)

These details are included to provide an idea of the scope of such experiments, and to show the development of the methods scientists are beginning to develop to measure how efficacious cyber defenses actually are. Provided enough tools are included in these large-scale tests, the state of the art as used in practice emerges. For the first time, consumers can see gaps in defenses across all choices, identify combinations of tools that maximally enhance defense (known as "defense in depth"), and researchers/technologists can prioritize next-step efforts. Importantly, policies mandating such testing will balance and scrutinize the marketing promises, often stated as a readily offered panacea for otherwise hard problems.

Step three, which entails a pilot study of tools in real-world conditions but with limited scope, has yet to be completed by our team. Such a study for detection and prevention tools would necessitate deploying each network defense tool in a network along with "honeypots"—computers donning realistic configurations but lacking in defense mechanisms—which would function as the analogous clinical trial "placebo." This study allows for direct comparison between control and experimental results (e.g., statistics can be computed for comparing the health of real network resources compared to their honeypot counterparts). Beyond tool efficacy, analyst usability experience measures should also be considered in this stage, providing a new dimension to this means of testing.

Because our previous experience is in testing security software, monitoring how tools perform when widely deployed in the real world is a natural extension of this. Yet, it is unclear how exactly to measure widespread performance, as this also may necessitate information sharing among disparate network operators. The aversion of network operators to share data notwithstanding, many modern security tools used in practice rely on cloud connections for updates and data processing, especially for applying AI models. This opportunity provides telematics feedback from each network across the globe to the vendor-operated cloud servers. The data informs threat intelligence and can strengthen AI models via retraining. Furthermore, the security as a service industry is growing, allowing providers to gather information from a wide variety of deployed security tools. In summary, private markets are leveraging opportunities to learn from widespread network defense tools in the wild, and the crowdsourced intelligence generated for consumers (e.g., network security professionals) is evidently worth the risk.

Notably, there are private and government efforts to catalogue software vulnerabilities, software platforms, types of weaknesses, and mitigations. This publicly available information has proven to be a great boon to the information security community and illustrates how all parties benefit from upkeep and organization of information from those willing to contribute. Upon initial disclosure of a vulnerability, a structured entry appears in the Common Vulnerabilities and Enumerations (CVE),[4] a publicly available index of software vulnerabilities hosted by MITRE, a private corporation. After appearing in the CVE dictionary, each vulnerability is ingested and enriched with much more information by the National Vulnerability Database (NVD),[5] hosted and actively managed by the US National Institute for Standards and Technology. The NVD includes cross-references to other related information also held in enumerated standards (e.g., referencing the appropriate entry of the Common Weakness Enumeration (a dictionary of types of insecurities in software and hardware) and of affected software in the Common Platform Enumeration (a dictionary of software including vendors and versions). As the name suggests, CVE is the de facto standard for cataloguing vulnerabilities; for instance, security professionals, the NVD, and other databases refer to vulnerabilities by their CVE number. In addition to these government-assisted sources, many other public vulnerability databases exist (e.g, [https://resources.whitesourcesoftware.com/blog-whitesource/open-source-vulnerability-databases](https://resources.whitesourcesoftware.com/blog-whitesource/open-source-vulnerability-databases)). In parallel, as vulnerabilities and correlated information are being catalogued, vendors of the affected software are, in general, creating mitigations (e.g., software updates and patches), and posting in their public-facing bulletins, referencing the known vulnerabilities alongside their patches. Overall, through private and public efforts to continually catalogue, update, and make available well-organized and accurate information on security vulnerabilities and patches, a system and an infrastructure has emerged to quickly index and share security problems and their fixes. This provides a wealth of needed resources for security operations to protect networks.

---

[4] https://cve.mitre.org/.
[5] https://nvd.nist.gov/vuln.

# Discussion

In an article discussing the regulation of AI-based systems that replace humans in decision making processes, Mulligan et al. advocate "a move from a procurement mindset to a policymaking mindset" (2019). While the motivation and application differ in cybersecurity, we also advocate a move from procuring the best existing AI-based defensive cybertechnologies to establishing policies that govern the entire life cycle of such technologies. These policies would be intended to establish accountability and transparency, which we believe would result in more effective AI-based tools for cybersecurity. As noted by Wieringa in a systematic survey of literature on algorithmic accountability, accountability "both exposes issues and promotes better behavior because people know they are being watched" (2020).

In our own experience testing ML-based tools for the US navy, we saw firsthand how rigorous testing and feedback can both enhance existing commercial tools (by providing feedback to the vendors) and aid in effective procurement. We suggest that the four-phase clinical trial process can be mapped and applied to the development of AI-based tools to provide both accountability and transparency. While we make no specific claims about the most effective way to implement this process via policy, in this section we discuss potential actions policymakers could take to implement each of these four phases and the limitations of our approach.

## Phase 1

Phase 1 requires domain experts to evaluate a technology's methodology to determine soundness. While our experiences documented above discuss third-party consultants meeting with vendors for this purpose, a more scalable option is already in place for cybersecurity. Specifically, Gartner

([https://www.gartner.com/en/information-technology](https://www.gartner.com/en/information-technology)) provides market research, in particular for cybersecurity submarkets, with a goal of providing information on the value of particular vendors' solutions. In addition, Mitre Att&ck Evaluations ([https://attackevals.mitre-engenuity.org/](https://attackevals.mitre-engenuity.org/)) have entered the cybersecurity market research sector, with a goal of "enabling users to better understand and defend against known adversary behaviors through a transparent evaluation process and publicly available results." Both provide information on the pros and cons of cybersecurity technologies without rigorous testing. In short, industry is providing a Phase 1 analogue, and the challenge before us is to provide, either through policy or some other means, the credibility and integrity of these sources.

Phase 2

Multiple research efforts besides ours are creating benchmarking abilities (e.g., datasets, evaluation methodologies), and/or novel metrics and measures for AI and cybersecurity technologies (ATT&CK evaluations, 2021; Myneni, 2020; Ring, 2019; Shah, 2020), yet most such efforts conclude with an academic publication. Implementation of an AI and/or cyber-focused clinical trial will require strategies for determining which evaluation methods are merit-worthy, ideally automating them, and finally incorporating them into a regular process with publicly available results. What is needed are clearly defined and transparent criteria, approaches, and requirements for testing and evaluating the resilience and reliability of AI-based methods for cyber defense. The government can play the role of a trusted third party and an "honest broker" that can facilitate testing and evaluation of the AI-based cyber defense approaches supplied by commercial entities and private industry. In this role, the government can set standards for testing and evaluation, host "bake-off" events, and impartially track and evaluate results while remaining completely impartial and protecting the intellectual property of the participants.

      As a thought experiment, suppose a national or international body curated, regularly updated, and publicly facilitated a test for some subset of cybersecurity tools, similar to Kaggle ([www.kaggle.com](www.kaggle.com)) competitions.[6] The idea is that, for a specific class of cybertechnologies leveraging AI/ML, one can submit a tool and be provided with an evaluation automatically. Similar to R-value insulation ratings, food labels, and vaccine effectiveness statistics, transparency will benefit consumers, establish the state of the art, and even possibly promote healthy competition for vendors. It is critical that testing methodology maintain both impartiality and integrity. Thus, it is imperative that third-party expertise remain a requirement for both administering and changing tests. It is also worth noting that deprecated test datasets may be made public for cybersecurity researchers. Finally, novel and varied tests may comprise adversarial AI techniques so as to provide a stress test of cyber tool resilience to these techniques. Overall, we believe such testing

---

[6] Kaggle designs and hosts ML competitions often by clearly and technically defining a desired task, making public a training dataset, and holding secret a testing dataset for evaluation of submitted technologies.

infrastructure, shared datasets, and the resulting transparency will enhance the development of tools in addition to providing needed information to the public.

Notably, existing initiatives can help evaluate the security and viability of existing AI models. One such initiative is the GARD program, orchestrated by the Defense Advanced Research Programs Agency (DARPA)[7]. The program is intended to create a platform for the evaluation of attacks and defenses against adversarial AI. This initiative presents one possible way for purposeful and principled evaluation of AI models against adversarial exploitation. Furthermore, it is a model where the government serves as a host and convener of the approaches developed and executed by the academic and commercial entities.

Phase 3

Similar to vaccine trials, pilot testing of AI and/or cybersecurity defense tools in a real network can provide insights into their effectiveness. For the analogue of placebos, we propose honeypots—computer systems donning no defenses and possibly with known vulnerabilities present—with the goal of monitoring adversarial techniques. Ideally, honeypots will be instantiated alongside close-to-identical workstations and servers, the latter employing defensive measures under test. Overtime, monitoring of defended attacks on the actual workstations versus the undefended activity on the honeypots will provide measures of the efficacy of defensive techniques. This whole process can be expedited by red team events, where hired offensive computing professionals wage cyberattacks to identify weaknesses in the defenses of the network and/or confuse or bypass AI models. Research for creating, tuning, and operationalizing such a process would be the next step to pursue such evaluations.

Phase 4

In the final phase, deployed tool monitoring provides a means of both recalling ineffective tools as well as retraining models to address attack vectors on the underlying algorithms as they arise. Having a public database of exploits against different classes of algorithms and methods for protecting an AI-based system against these exploits, including a way to access data sets to harden models if required, would be a good place to start. This database could also include attacks against specific tools, whether or not the tool vendor has provided sufficient protection against the attack, the release number in which the patch occurred, and a recommended course of action for organizations considering using that tool on their network. The CVE and NVD databases are used to track software vulnerabilities and offer a model for what this process might look like in practice. In addition, MITRE has a well-defined process for how to report such vulnerabilities[8]. Because vulnerabilities in AI systems are fundamentally different from software vulnerabilities, the structure and content of such a database would look differently, but a similar reporting process could be used.

---

[7] https://www.darpa.mil/program/guaranteeing-ai-robustness-against-deception.
[8] https://cve.mitre.org/cve/researcher_reservation_guidelines.

To ensure that AI-based tools are properly secured in operational environments, it would also be helpful to establish clear policies that security operations centers (SOCs) can follow to ensure a secure deployment as well as a pathway allowing SOCs to provide feedback that can be used to improve the tools themselves. Examples of helpful policies include guidelines for protecting training data that could be used or poisoned by an adversary and steps for integrating these tools effectively into existing infrastructure. Regarding feedback from SOCs, Engstrom and Ho (2020) propose that having random subsampling of cases quality checked by a human operator will provide a more tractable measure of how an automated mechanism is performing in practice. Given that SOCs often perform manual investigation of logs and incidents, they are ideally situated to provide this type of feedback. Findings of investigations are usually well documented by ticketing systems and preserved by saving correlated network data in the tickets. Research and mechanisms for "closing the loop," (i.e., leveraging these ongoing manual efforts to systematically find and fix problems with automated tools) is burgeoning (e.g., see Veeramachaneni et al. [2016]), but no widespread practices have been established. Creating a clear pipeline to receive feedback on existing deployments from SOCs, addressing any identified issues, and then pushing resultant updates out to all SOCs using that tool would be extremely beneficial.

Recalling products, while used in many other sectors, is difficult for cybersecurity technologies because organizations may be locked into a contract with a vendor and susceptible to the sunk cost fallacy. It is also important to avoid giving one vendor a monopoly or discouraging vendors from innovating by making it difficult to turn a profit. It may be that market forces and self-interest are sufficient in most cases as long as transparency into tool efficacy is encouraged and maintained. More research is needed to determine when ad if forced recall of AI-based tools makes sense and how that process should be governed.

## Limitations

Because the resilience of AI-based cyber defense tools depends on the robustness of the underlying algorithms and ensuring the robustness of such algorithms is an area of open research (Carlini, 2020; Goodfellow & Papernot, 2017), no vetting process can protect against every potential exploit because the space of potential exploits has not been enumerated. As noted by Carlini, "It's very difficult to differentiate between true robustness and apparent robustness, where true robustness means that no adversarial example actually exists, and apparent robustness means it looks that way because we haven't found one yet" (2020). However, that does not by any means suggest that vetting is useless.

Many of the cryptography algorithms used millions of times every single day are not known to be one hundred percent secure. There might be a vulnerability that has not yet been discovered or at least it is not publicly known. Furthermore, because these cryptography algorithms go through a rigorous and transparent vetting process, they effectively protect private transactions and data from most types of adversaries. Similarly, rigorous testing of AI-based cyber defense tools will go a long way in making our networks more secure, even if it doesn't ensure protection against every edge case or zero-day exploit.

# Conclusion

The analogies, structures, and processes we present in this article are ultimately an exercise in formalization of quality assurance and testing practices inspired by another domain. In our case, that other domain is pharmacology and immunology. As with any other inspiration, it is simply that: an inspiration. We do not place any strong claim that the development and evaluation of new vaccines is an exact solution or a precise model for the problems of cyber defense reliability and resilience. However, the problems we highlight—namely, the critical need for the use of AI in cyber-defense, and the inherent uncertainty about its ultimate resilience to adversarial exploitation—are real problems that need to be addressed. Furthermore, our early experiences as well as the rigor and discipline that we apply in this problem domain are valuable and worth formalizing. Our attempt to map our process to the process of drug discovery and vaccine safety is to propose a practice with even greater rigor. For that reason, we hope to inspire policymakers to observe some of the practices we propose and consider them as foundations for a future, formal process. The cyberthreat and the risks of AI adversarial exploitation are real, and so is the need for a rigorous, transparent, disciplined, and staged evaluation of AI-based cyber defense tools.

# Acknowledgments

We wish to thank reviewers that have helped polish this document. The research is based upon work supported by the US Department of Defense (DOD), Naval Information Warfare Systems Command (NAVWAR), via the US Department of Energy under contract DE-AC05-00OR22725. The views and conclusions contained herein are those of the authors and should not be interpreted as representing the official policies or endorsements, either expressed or implied, of the DOD, NAVWAR, or the US Government. The US Government is authorized to reproduce and distribute reprints for governmental purposes notwithstanding any copyright annotation thereon.

McDermott, K. (2020, December 7). Winners of artificial intelligence challenge announced. Retrieved January 31, 2021, from https://www.navy.mil/Press-Office/News-Stories/Article/2436651/winners-of-artificial-intelligence-challenge-announced/.

MacDonald, N., et al. (2020). Global vaccine action plan lessons learned I: Recommendations for the next decade." *Vaccine 38*(33), 5364–5371. https://doi.org/10.1016/j.vaccine.2020.05.003.

Mulligan, D. K., and Bamberger, K. A. (2019). Procurement as policy: Administrative process for machine learning. *Berkeley Technology Law Journal 34*, 781–858. http://dx.doi.org/10.2139/ssrn.3464203.

Myneni, S., Chowdhary, A., Sabur, A., Sengupta, S., Agrawal, G., Huang, D., & Kang, M. (2020, August). Dapt 2020-constructing a benchmark dataset for advanced persistent threats. In *International Workshop on Deployable Machine Learning for Security Defense* (pp. 138-163). Springer, Cham.

Naval Information Warfare Systems Command (NAVWARSYS). US Department of Defense. AI ATAC 3 challenge: Efficiency & effectiveness afforded by security orchestration & automated response (SOAR) capabilities. Challenge website. https://www.challenge.gov/challenge/AI-ATAC-3-challenge/.

NAVWARSYS. (n.d.-b). Artificial intelligence applications to autonomous cybersecurity (AI ATAC challenge. Challenge website. https://www.challenge.gov/challenge/artificial-intelligence-applications-to-autonomous-cybersecurity-challenge/.

NAVWARSYS. (n.d.-c). US Department of Defense. Network detection of adversarial campaigns using artificial intelligence and machine learning. Challenge website. https://www.challenge.gov/challenge/network-detection-of-adversarial-campaigns/.

Schneier, B. (2021, January 5). The SolarWinds hack is stunning. Here's what should be done. Schneier on Security, retrieved from www.schneier.com/essays/archives/2021/01/the-solarwinds-hack-is-stunning-heres-what-should-be-done.html.

Ring, M., Wunderlich, S., Scheuring, D., Landes, D., & Hotho, A. (2019). A survey of network-based intrusion detection data sets. *Computers & Security*, *86*, 147-167.

Shah, N., Ho, G., Schweighauser, M., Ibrahim, M., Cidon, A., & Wagner, D. (2020, August). A Large-Scale Analysis of Attacker Activity in Compromised Enterprise Accounts. In *International Workshop on Deployable Machine Learning for Security Defense* (pp. 3-27). Springer, Cham.
14

## Author Biographies

Edmon Begoli, PhD, is a distinguished member of the research staff with the Oak Ridge National Laboratory (ORNL) and is the Head of the Adversary Intelligence Systems Section. His research interests are in resilient and reliable system architectures for large-scale data analysis in mission critical domains. Edmon holds undergraduate, graduate, and doctoral degrees in Computer Science, and is an adjunct professor of Computer Science at the University of Tennessee-Knoxville, EECS department.

Kathryn Knight is a member of the Data Lifecycle and Scalable Workflows group within the National Center for Computational Sciences at Oak Ridge National Laboratory. She is also a Ph.D. candidate in the School of Information Sciences at the University of Tennessee. Her research interests revolve around information architecture and knowledge management, particularly related to how information is organized, represented, retrieved, and used.

Sean Oesch, PhD is a software architect turned security researcher. His recent research focuses on applications of ML to cyber and usable security. While at ORNL he has worked on a wide variety of projects, from cyber forensics to using ML to optimize fuel efficiency at stoplights. He completed his PhD at University of Tennessee, Knoxville.

Robert A. (Bobby) Bridges received a B.S. in 2005 from Creighton University and a Ph.D. in 2012 from Purdue University, both in Mathematics. Bobby joined Oak Ridge National Laboratory as a postdoc in 2012, was promoted to Associate Researcher in 2013 then Staff Researcher in 2017, and now serves as the Acting Cybersecurity Group Leader. Bobby currently leads a team providing ongoing scientific support to assist the US Navy in acquisition decisions through large-scale, scientific testing of market-leading cybersecurity tools.